\documentclass[letterpaper,journal]{IEEEtran}

\newtheorem{remark}{Remark}
\usepackage{amssymb}
\usepackage{amsmath,amsfonts}
\usepackage{mathtools}
\usepackage{newtxtext} 
\usepackage[hidelinks]{hyperref}
\usepackage{booktabs}
\usepackage{threeparttable}
\usepackage{tabularx}
\usepackage{xcolor}

\usepackage[backend=biber,style=ieee,natbib=true,maxcitenames=2,mincitenames=1]{biblatex}

\newcommand{\bfm}[1]             { \mathbf{#1}     } 
\newcommand{\Nabla}       { \boldsymbol{\nabla} }   
\newcommand{\SLOinf}            { L^\infty(\Omega)}          
\newcommand{\SLTO}            { L^2(\Omega) }

\newcommand{\MSGE}            { \operatorname{MSGE} }
\newcommand{\MAXI}            { \operatorname{MAXI} }
\newcommand{\MAXG}            { \operatorname{MAXG} }
\newcommand{\GSNR}            { \operatorname{GSNR} }

\newcommand{\DD}            { \kappa  }              
\newcommand{\bb}             { {\bfm{b}}   }             
\newcommand{\vn}             { \bfm{n}     }             

\newcommand{\e}{\text{e}}

\usepackage{graphicx} 

\usepackage{fancyhdr}
\begin{document}

\title{PC-SRGAN: Physically Consistent Super-Resolution Generative Adversarial Network for General Transient Simulations}

\author{Md Rakibul Hasan, Pouria Behnoudfar, Dan MacKinlay, Thomas Poulet%
\thanks{This work was supported by resources provided by the Pawsey Supercomputing Research Centre with funding from the Australian Government and the Government of Western Australia.}
\thanks{We also acknowledge the use of the CSIRO Bracewell computing cluster.}
\thanks{M R Hasan, T Poulet, and P Behnoudfar (during the study) are with the Commonwealth Scientific and Industrial Research Organisation (CSIRO) Mineral Resources, Kensington, WA 6151, Australia.}
\thanks{P Behnoudfar is with the Department of Mathematics, University of Wisconsin-Madison, Madison, WI 53706, USA.}
\thanks{D MacKinlay is with the Commonwealth Scientific and Industrial Research Organisation (CSIRO) Data61, Eveleigh, NSW 2015, Australia.}
\thanks{M R Hasan is also with the School of Electrical Engineering, Computing and Mathematical Sciences, Curtin University, Bentley, WA 6102, Australia.}
\thanks{M R Hasan is also with the Department of Electrical and Electronic Engineering, BRAC University, Dhaka 1212, Bangladesh.}

\thanks{Corresponding author: Pouria Behnoudfar (email: behnoudfar@wisc.edu)}
}

\maketitle
\thispagestyle{fancy}

\begin{abstract}
Machine Learning, particularly Generative Adversarial Networks (GANs), has revolutionised Super-Resolution (SR). However, generated images often lack physical meaningfulness, which is essential for scientific applications. Our approach, PC-SRGAN, enhances image resolution while ensuring physical consistency for interpretable simulations. PC-SRGAN significantly improves both the Peak Signal-to-Noise Ratio and the Structural Similarity Index Measure compared to conventional SR methods, even with limited training data (e.g., only 13\% of training data is required to achieve performance similar to SRGAN). Beyond SR, PC-SRGAN augments physically meaningful machine learning, incorporating numerically justified time integrators and advanced quality metrics. These advancements promise reliable and causal machine-learning models in scientific domains. A significant advantage of PC-SRGAN over conventional SR techniques is its physical consistency, which makes it a viable surrogate model for time-dependent problems. PC-SRGAN advances scientific machine learning by improving accuracy and efficiency, enhancing process understanding, and broadening applications to scientific research. We publicly release the complete source code of PC-SRGAN and all experiments at \url{https://github.com/hasan-rakibul/PC-SRGAN}.
\end{abstract}

\begin{IEEEkeywords}
Physics-informed neural network, physical consistency, super-resolution, generative adversarial network, image quality assessment, time integrator.
\end{IEEEkeywords}

\section{Introduction}

\begin{figure}[t!]
\centering
\includegraphics[width=1\linewidth]{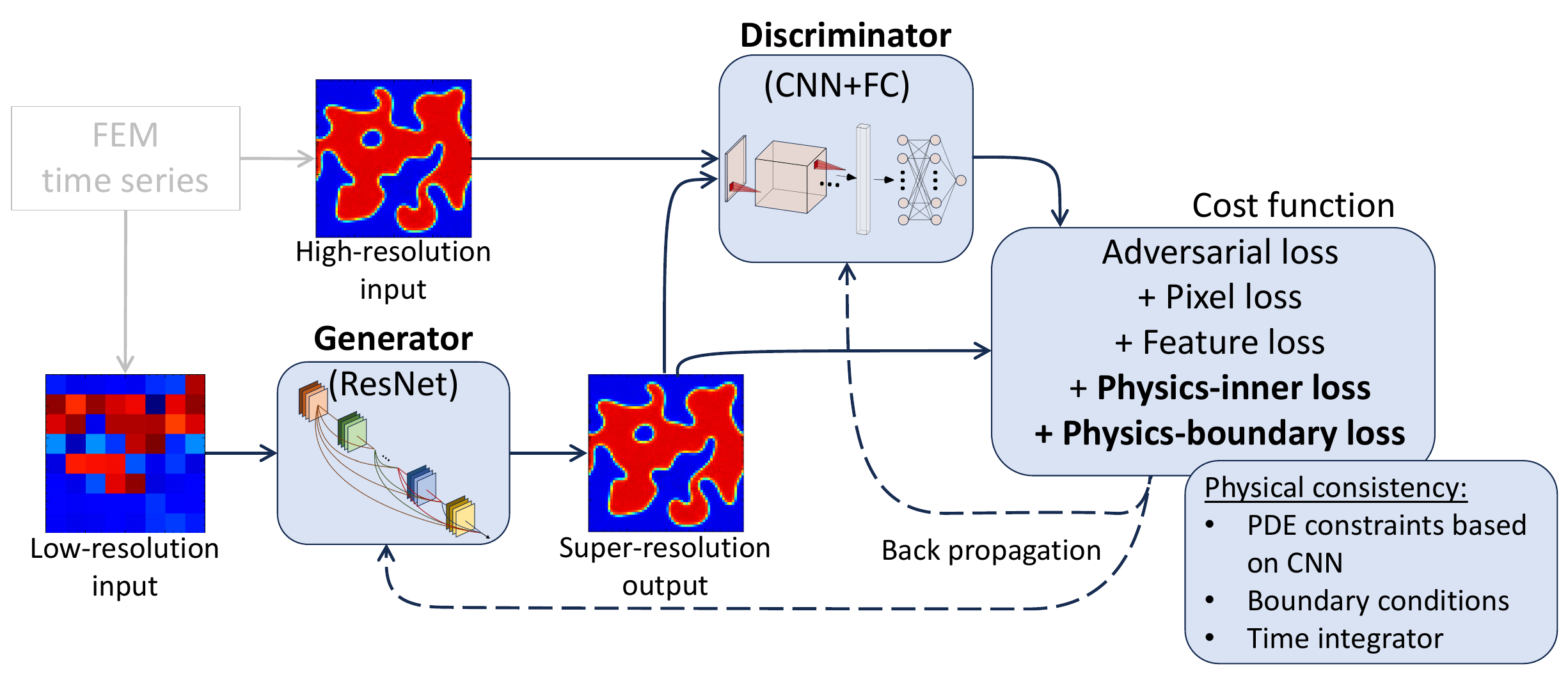}
\caption{Schematic description of PC-SRGAN. A Residual Network (ResNet)-based Generator model generates each high-resolution time step from the corresponding low-resolution simulation result. The Discriminator model, constructed by a Convolutional Neural Network (CNN) and a Fully Connected (FC) network, judges the generated image based on the high-resolution ground truth. As for the training using the backpropagation algorithm, we introduce physics-based losses into the traditional SRGAN~\citep{ledig2017photo} losses.}
\label{fig:method_schematic}
\end{figure}

\IEEEPARstart{S}{uper} Resolution (SR) infers high-resolution images consistent with low-resolution representations. Among its numerous applications, SR can be used to dramatically reduce the computational expense of numerical simulations for physical processes, such as transient fluid flows, where high-resolution outputs are typically expensive to compute directly.

SR is a particularly active research area, as demonstrated by a rich taxonomy of techniques. The review of \citet{LepchaImage2023} proposes a division of the four main categories of methods: interpolation, reconstruction/regularisation, predictive learning and transformers~\citep{LepchaImage2023}. They all present both advantages and shortcomings. Predictive learning-based approaches can attain superior results in challenging scenarios where interpolation methods struggle, for example, in multi-modal or rough sample spaces. In turn, predictive learning SR encompasses several overlapping categories, including Convolutional Neural Networks (CNNs), Residual Networks (ResNets), Densely-connected Networks (DenseNets), diffusion models, and Generative Adversarial Networks (GANs). GANs show particular promise in SR, both in generating general synthetic images, video, or voice data that are perceptually close to real data, and for applications to numerical simulations in particular~\citep{bao2022physics,XieTempoGAN2018}. Unlike diffusion models, GANs have a highly efficient sampling process~\citep{NicholImproved2021} and potentially accelerate inference by allowing efficient high-resolution inference from cheaper, low-resolution samples.

GANs \citep{goodfellow2014generative} use two neural networks trained together in an adversarial process. The two networks are the generator model, which attempts to generate realistic synthetic data, and the discriminator model, which attempts to differentiate between real and synthetic data. This competition between the two drives the generator to produce increasingly realistic samples, while the discriminator becomes increasingly sophisticated at distinguishing fake from synthetic data.

Prior works on physics-guided approaches show promise in generating high-resolution simulations from low-resolution simulations. For instance, \citet{XieTempoGAN2018} introduced tempoGAN for temporally consistent Super-Resolution (SR) of fluid flows. The generator is conditioned on physical quantities from low-resolution fluid data like velocities and vorticities, which enables it to infer more realistic high-resolution details. A physics-aware data augmentation technique is also introduced to reduce overfitting and memory requirements during training. Together, these contributions result in a model that can instantly generate fluid flows with highly detailed, realistic textures and coherence across frames using just a single low-resolution time step as input. Those results, however, did not exploit physical consistency constraints other than through the training data. Subsequently, \citet{bao2022physics} proposed a Physics-Informed Neural Network (PINN) for reconstructing high-resolution Direct Numerical Simulation (DNS) from low-resolution Large Eddy Simulation (LES) of turbulent flows. That method incorporates a physics-guided SR model by imposing a physical constraint -- the divergence-free property of incompressible flow; yet, this constraint is specific to the selected problem and only enforces part of the underlying physics simulated. The same approach was also used with GANs \cite{Bode2021}. However, the mentioned techniques are inherently flawed; they display a major causality drawback due to first-order time derivatives in such scenarios. Indeed, their training methodologies incorporate information from subsequent temporal data points to derive solutions for the current time. Yet, the solution at any specific point in time must solely depend on the initial conditions, boundary conditions, and previous states~\citep{grigis1994microlocal}.

In this contribution, we propose a physics-based learning method \citep{KadambiIncorporating2023} to account explicitly for the governing equations of the physical problems and improve the quality of the generated SR output through physical consistency. We target general transient Initial-Boundary-Value Problems (IBVPs) with first-order time derivatives and select two standard specific problem types to demonstrate the approach (see \autoref{sec:method}), particularly prevalent for transient fluid flow simulations. We use an external Finite Element Method (FEM) solver to generate corresponding numerical simulation results in two dimensions on two separate regular grid sizes, to produce images of different resolutions with two scaling factors ($8\times$ and $4\times$) along each axis. We then modify the SRGAN approach by adding two novel physics-based loss components -- one for inner pixels and another for boundary pixels -- to ensure physical consistency (\autoref{fig:method_schematic}). That physical loss covers the constraints from the Partial Differential Equation (PDE) on the specified domain, along with corresponding boundary conditions and time integrator for the discrete time-marching strategy.

\section{Method}\label{sec:method}
Our approach integrates a physics-based loss function into a Super-Resolution GAN (SRGAN) to ensure physical consistency in the high-resolution output. We target general transient initial-boundary-value problems with first-order time derivatives and demonstrate our method using two specific problems: the Eriksson-Johnson and the Allen-Cahn equations. We refer interested readers to Appendix A for a detailed description of these two examples and their real-world applications.

\subsection{Improved super-resolution accuracy by including physics}

We assume $\Omega \subset \mathbb{R}^N$, $N \geq 1$ is an open bounded domain with Lipschitz boundary $\partial \Omega$ and $T$ is the final time. Then, we define $\Omega_T = \Omega \times (0, T)$ to be the space-time domain which is open and bounded with a  Lipschitz boundary $\partial \Omega_T = \partial \Omega \cup \Gamma_{0} $, where $\partial \Omega = \Gamma_{in} \cup \Gamma_{out}$ with $\Gamma_{in}$ and $\Gamma_{out}$ being the in- and out-flow boundaries, respectively, and $\Gamma_{0}$ is the initial-time condition. This allows us to consider a general transient Initial-Boundary-Value Problem (IBVP) with a first-order derivative in time as:
\begin{equation}\label{eq:prob}  
\frac{\partial u }{\partial t}  +\mathcal{F}(u) =0,\quad \text{ in } \Omega_T.  
\end{equation}

The functional $\mathcal{F}(u)$ encapsulates spatial operators (e.g., gradient), a source function, nonlinear operators, physical properties, and boundary conditions. Initial and boundary data on $\Gamma_{0}$ and $\partial \Omega$, respectively, are given such that~\eqref{eq:prob} admits a well-posed equation. 

\subsubsection{Dataset}\label{subsubsec:data}
Our proposed method is designed based on the general form~\eqref{eq:prob} and can estimate any problems with such a form. In the sequel, we extend our technique to different boundary conditions: Dirichlet, Neumann, and Periodic conditions.

First, using the method of lines, we discretise~\eqref{eq:prob} using FEM in the spatial domain and a time-marching technique to obtain the fully discrete problem. Then, we generate low- and high-resolution (ground truth) time series, with a specific resolution increase in each dimension between the resolutions of the series. Across different setups, we increase the resolution 8 times: $8\times8$ images to $64\times64$ images. This results in a dataset for specific physical parameters containing a time series of solutions satisfying~\eqref{eq:prob} at the initial time step until the last. Therefore, the content of each image is a snapshot of fluid flow at different time steps. For the Allen-Cahn problem with the Periodic boundary condition, we further generate data with $4\times$ resolution increase (i.e., $8\times 8 \rightarrow 32\times 32$).

Note that image SR methods, such as SRGAN \citep{ledig2017photo} and ESRGAN \citep{esrgan}, which are designed for natural images, create low-resolution images by downsampling high-resolution ones. In contrast, we generate both low- and high-resolution snapshots independently using the FEM simulation with different mesh sizes, thereby resonating with our broader application of reconstructing fine-resolution solutions from coarse simulations. Refer to Appendix Table I for the number of samples in training, validation, and test splits in all four datasets used in this study. We release all datasets publicly at \citep{hasan2025transient}.

\subsubsection{Loss function}
Our Physically Consistent SRGAN (PC-SRGAN) builds upon SRGAN \citep{ledig2017photo}, a generative adversarial network designed for Super Resolution (SR) tasks, which is proficient in generating lifelike $4\times$ upscaled images across a two-dimensional space. A summary of the SRGAN architecture is provided in Appendix D.

Many SR methods measure resolution quality using a pixel loss, most commonly the pixel-wise MSE between the generated high-resolution image and the ground truth high-resolution image~\citep{LepchaImage2023}:
\begin{equation}
\mathcal{L}_\text{px} = \text{MSE}(U_\text{sr}, U_\text{gt}),
\end{equation}
where $U_\text{sr}$ and $U_\text{gt}$ denote the generated super-resolved and ground truth images, respectively. Reliance on such pixel-wise objective alone results in suboptimal perceptual quality of the generated image \citep{ledig2017photo}. SRGAN \citep{ledig2017photo} circumvents this weakness by using a weighted sum of content loss (also termed as ``feature loss'') and adversarial loss. The content loss is derived from the VGG19 \citep{simonyan2015deep} feature representations of the generated and ground truth images, calculated as the MSE between their ReLU activations. The adversarial loss is a Binary Cross-Entropy (BCE) loss that measures the discrepancy between classifications assigned to the real and generated samples. This loss component encourages the generator to produce high-resolution images that can potentially fool the discriminator $D$. Overall, the training objective in SRGAN can be written as:
\begin{equation}
\begin{split}
\mathcal{L}_\text{SRGAN} &= w_2 \cdot \text{MSE}\left(\psi_{V}(U_\text{sr}\right), \psi_{V}\left(U_\text{gt})\right) \\
& + w_3 \cdot \text{BCE}\left(D(U_\text{sr}), D(U_\text{gt})\right)
\end{split}
\end{equation}
where $\psi_{V}$ is the VGG19 feature representation. $w_2$ and $w_3$ are tunable weights for the corresponding loss components.

In PC-SRGAN, we incorporate two additional physics-based terms (\autoref{fig:method_schematic}) to ensure physical consistency of the solutions, i.e., to encourage the generated high-resolution images towards the underlying physical principles. These terms penalise deviation from valid solutions specified by the PDE within the domain and under given boundary conditions. We keep the pixel loss in the final cost function to enable ablation studies of individual components.
\begin{equation}
\mathcal{L}_\text{PC-SRGAN} = w_1 \cdot \mathcal{L}_\text{px} + \mathcal{L}_\text{SRGAN} + w_4 \cdot \mathcal{L}_{\phi}^{\text{in}} + w_5 \cdot \mathcal{L}_{\phi}^{\text{bd}}
\label{eq:loss_pc_srgan}
\end{equation}
where $w_1$, $w_4$ and $w_5$ are tunable weights for corresponding loss components. We formulate these physics-based terms based on a semi-discrete version of equation~\eqref{eq:prob}, achieved through a time-marching methodology. The inner (i.e., excluding boundary pixels) physics loss for different time integrators is defined as:
\begin{multline}
\mathcal{L}_{\phi}^{\text{in}} = \\
\begin{dcases}
\frac{3}{2\tau}(U_\text{sr}^n - \frac{4}{3}U_\text{gt}^{n-1} + \frac{1}{3}U_\text{gt}^{n-2}) + f(U_\text{sr}^n, \epsilon, K, r, \theta),  \text{BDF,}\\
\frac{1}{\tau}(U_\text{sr}^n - U_\text{gt}^{n-1}) + \frac{1}{2}(f(U_\text{sr}^n, \epsilon, K, r, \theta) \\+ f(U_\text{gt}^{n-1}, \epsilon, K, r, \theta)),  \text{CN,}\\
\frac{1}{\tau}(U_\text{sr}^n - U_\text{gt}^{n-1}) + f(U_\text{gt}^{n-1}, \epsilon, K, r, \theta),  \text{EE,}
\end{dcases}
\end{multline}
where $f(\cdots)$ is the spatial operator, depending on the problem:

\begin{multline}
    f(u, \epsilon, K, r, \theta) = \\
    \begin{dcases}
        -\epsilon \Nabla u + \frac{0.5}{\mathcal{E}^2}K \theta \log\frac{1+u}{1-u}-1.2u,  \quad \text{Allen-Cahn,} \\
        -\epsilon \Nabla u + r\cos{\theta}\frac{\partial u}{\partial x} + r\sin{\theta}\frac{\partial u}{\partial y} \\+ Ku(u-1),  \quad \text{Eriksson-Johnson,}
    \end{dcases}
    \label{eq:spatial_op}
\end{multline}
where $\epsilon, K, r$ and $\theta$ are FEM parameters used to generate corresponding $u$. For the Allen-Cahn problem, the constant $\mathcal{E}$ is defined as: 
\begin{equation}
\mathcal{E} = \frac{1}{64} \times \frac{1}{2\times(2^{0.5})\times \tanh{(0.9)}}
\end{equation}
The boundary losses are calculated using MSE, with the option to apply Dirichlet, Periodic or Neumann boundary conditions:
\begin{equation}
    \mathcal{L}_{\phi}^{\text{bd}} = 
    \begin{cases}
       \sum_\omega^{l,r,t,b}\text{MSE}(U_{\text{sr}}^\omega, U_{\text{gt}}^\omega) , & \text{Dirichlet} \\
       \text{MSE}(U_{sr}^l, U_{sr}^r) + \text{MSE}(U_{\text{sr}}^t, U_{\text{sr}}^b), & \text{Periodic} \\
       \sum_\omega^{l,r,t,b}\text{MSE}(U_{\text{sr}}^\omega(x), U_{\text{sr}}^\omega(x \pm h)) , & \text{Neumann}
    \end{cases}
\end{equation}
where $l,r,t$, and $b$ represent the left, right, top, and bottom boundaries, respectively. For the Neumann boundary condition, $U_\text{sr}(x)$ denotes the first pixel from the corresponding boundary (i.e., leftmost, rightmost, topmost or bottommost), and $U_\text{sr}(x \pm h)$ refers to the points adjacent to the corresponding boundary.

\subsection{Convolution filters for calculating spatial operators}

To compute second-order derivatives, we use a convolution-filtering kernel. This process is designed to prevent the enforcement of additional regularity to the parameter space compared to the classical physics-informed machine learning methods~\citep{cai2021physics,cai2021physicsf}. For example, automatic derivation in~\citep{raissi2019physics} results in approximated results with the regularity of $H^2$, limiting the performance for realistic problems with localised behaviours. Also, defining the directional derivatives allows us to account for heterogeneous diffusion coefficients $\epsilon $ as we can estimate $\nabla \cdot \epsilon \nabla $. We report the filters in Appendix B. 

\section{Results}\label{sec:results}

\begin{table*}[t!]
    \centering
    \caption{Comparison of SRGAN \citep{ledig2017photo}, ESRGAN \citep{esrgan} and our proposed PC-SRGAN on hold-out test sets. Overall, PC-SRGAN and PC-ESRGAN significantly outperform SRGAN and ESRGAN approaches by large margins across different experimental conditions. Among different time integrators, the Backward Differentiation Formula (BDF) results in the best performance compared to Crank-Nicolson (CN) and Explicit Euler (EE). \textbf{Boldface} indicates the best performance in each setup.}
    \label{tab:results}
    \begin{tabular}{@{}*5l*6c@{}}\toprule
        Scale factor & Data & Boundary type & Model & Time integrator & PSNR $\uparrow$ & SSIM $\uparrow$ & MSE $\downarrow$ & MSGE $\downarrow$ & LPIPS $\downarrow$ \\ \midrule
        $8\times$ & Allen-Cahn & Periodic & SRGAN \citep{ledig2017photo} & -- & 32.4400 & 0.9084 & 0.0111 & 18.1448 & 0.0272 \\ 
                &   &   & ESRGAN \citep{esrgan} & -- & 27.3812 & 0.8540 & 0.0237 & 33.0352 & 0.0521 \\ \cmidrule{4-10}
        &   &   & PC-SRGAN (Ours) & BDF & 37.2748 & 0.9652 & 0.0041 & 7.2372 & 0.0172  \\
            &   &   &   & CN & 35.0870 & 0.9582 & 0.0046 & 8.1293 & 0.0232 \\
            &   &   &   & EE & 33.9045 & 0.9499 & 0.0053 & 9.9475 & 0.0356 \\ \cmidrule{5-10}
            &   &   & \textbf{PC-ESRGAN (Ours)} & \textbf{BDF} & \textbf{44.2954} & \textbf{0.9792} & \textbf{0.0027} & \textbf{4.6497} & \textbf{0.0049} \\ \cmidrule{3-10}
            &   & Neumann & SRGAN \citep{ledig2017photo} & -- & 33.1739 & 0.9378 & 0.0045 & 8.5711 & 0.0190 \\
            &   &   & \textbf{PC-SRGAN (Ours)} & \textbf{BDF} & \textbf{38.1240} & \textbf{0.9761} & \textbf{0.0029} & \textbf{4.9526} & \textbf{0.0122} \\ \cmidrule{2-10}
        &   Eriksson-Johnson & Dirichlet & SRGAN \citep{ledig2017photo} & -- & 41.6019 & 0.9968 & 0.0005 & 0.0818 & 0.0014 \\
         &   &   & \textbf{PC-SRGAN (Ours)} & \textbf{BDF} & \textbf{49.3625} & \textbf{0.9995} & \textbf{0.0001} & \textbf{0.0124} & \textbf{0.0002} \\ \midrule
        $4\times$ & Allen-Cahn & Periodic & SRGAN \citep{ledig2017photo} & -- & 30.7071 & 0.9424 & 0.0061 & 11.3891 & 0.0046 \\
        &   &   & \textbf{PC-SRGAN (Ours)} & \textbf{BDF} & \textbf{33.1279} & \textbf{0.9737} & \textbf{0.0039} & \textbf{6.2260} & \textbf{0.0019} \\
        \bottomrule
    \end{tabular}
\end{table*}

\begin{figure}[t!]
\centering
\includegraphics[width=\columnwidth]{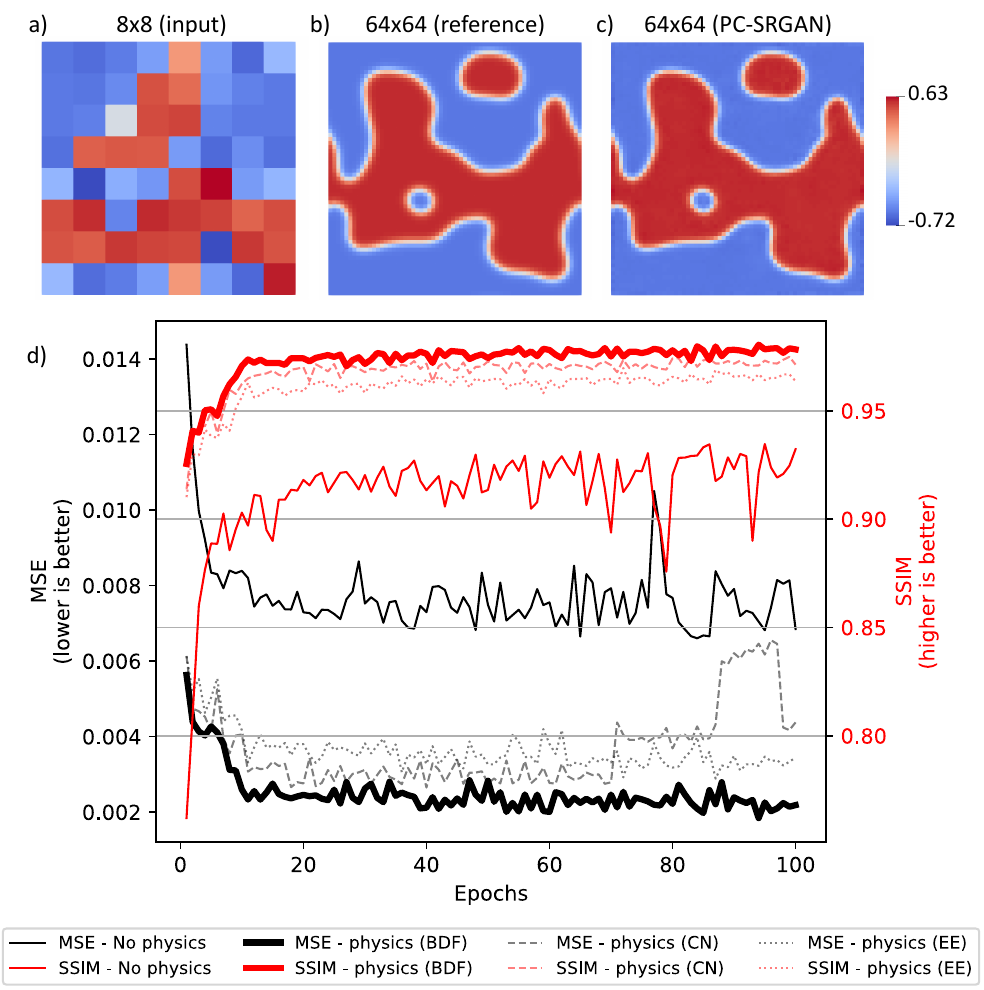}
\caption{(a) Example of low-resolution ground truth image, $8\times8$ grid; (b) high-resolution ground truth, $64\times64$ grid; (c) corresponding enhanced $64\times64$ result obtained with PC-SRGAN from the low-resolution one. 
(d) Evolution of Mean Square Error (MSE, lower is better) and Structural Similarity Index Measure (SSIM, higher is better) metrics with respect to training epochs, highlighting the significant improvement provided by the physical considerations, and comparing specifically the impact of three different time integrators: Explicit Euler (EE), Crank-Nicolson (CN) and Backward Differentiation Formula (BDF).}
\label{fig:SSIM}
\end{figure}

We conduct various tests for two common yet challenging physical systems, the Eriksson-Johnson and the Allen-Cahn equations. For each system, we compare the performance of PC-SRGAN against the baseline SRGAN~\citep{ledig2017photo}. We further analyse the effects of different time integrator algorithms on enforcing the dynamic evolution of the problem. This insight aims to enhance the effectiveness of existing methods (e.g., \citep{RenSuperBench2023}). The results in \autoref{tab:results} demonstrate improved performance across standard quality assessment metrics~\citep{SaraImage2019}, notably the Mean Square Error (MSE), Peak Signal to Noise Ratio (PSNR), Structural Similarity Index Measure (SSIM)~\citep{WangImage2004}, and Learned Perceptual Image Patch Similarity (LPIPS) \citep{zhang2018perceptual}. \autoref{fig:SSIM} shows detailed results for the Allen-Cahn system with Periodic boundary conditions. The evolution of the performance metrics as a function of the number of training epochs demonstrates that PC-SRGAN also accelerates convergence and reduces fluctuations during training.

The comparison between time integrators also reveals that a loss function based on multi-step methods, such as the Backward Differentiation Formula (BDF), produces superior outcomes compared to those derived from a multi-stage method -- e.g., the Crank-Nicolson (CN) approach categorised as a Runge-Kutta (RK) technique -- and the Explicit Euler (EE) formulation. \autoref{tab:results} concludes that for Allen-Cahn problems, using Periodic or Neumann boundary conditions, the PSNR increases by a large amount, nearly 5 dB by PC-SRGAN. Even for a smoother case of Eriksson-Johnson, the PSNR notably jumps by almost 8 dB.

We further experiment with Enhanced SRGAN (ESRGAN) \citep{esrgan} in one of the setups (BDF time integrator on the Allen-Cahn problem with the Periodic boundary conditions). ESRGAN consists of a CNN-based Residual-in-Residual Dense Block (RRDB) network as its generator. Following \citep{esrgan}, RRDB is first pre-trained separately and then integrated into ESRGAN for adversarial training. For a fair comparison, all models  -- i.e., SRGAN, ESRGAN, PC-SRGAN and PC-ESRGAN -- are trained using the same training samples. We found that ESRGAN overfits early during the training stage (best performance at epoch 17 out of 120), likely due to fine-tuning of its generator on the same training set. Between SRGAN and ESRGAN, the former is found to be better on our specific SR domain, which is notably different from the domain experimented with in the original ESRGAN work \citep{esrgan}. Such a better performance of SRGAN over ESRGAN is consistent with several works on image SR \citep{dai2021deep,deepthi2024novel}.

In contrast, our physics-based ESRGAN (i.e., PC-ESRGAN), which utilises the same pre-trained RRDB as its Generator, trains well (achieving its best results at epoch 107) and achieves the best SR metrics, while also benefiting from the visual quality promised by ESRGAN. PC-ESRGAN did not overfit like ESRGAN, presumably because the network learns physical consistency from the physics-based loss terms we impose during its training. 

The success of PC-ESRGAN suggests that our proposed method of imposing physical consistency works well on different GAN architectures. Since these physics-based loss terms are generic and model-agnostic, they may also apply to non-GAN SR models (e.g., diffusion- or flow-based methods), which we leave for future work.

Although SRGAN \citep{ledig2017photo} was originally demonstrated for $4\times$ upscaling, we evaluate our method on a more challenging $8\times$ upscaling across different dataset setups. For completeness, we also include results for the $4\times$ upscaling on the Allen-Cahn problem with the Periodic boundary condition. As shown in \autoref{tab:results}, our method outperforms the baselines at both $8\times$ and $4\times$ upscaling factors across all five evaluation metrics.

The scaling factor is not a fixed restriction of our method and could potentially be larger. Because PC-SRGAN adds physics-based constraints that guide the super-resolved outputs towards physically consistent solutions, it can be expected to maintain satisfactory performance even under more extreme upscaling factors, such as $16\times$. We leave this interesting direction for future work.

\subsection{Physical consistency and gradient fidelity metrics in evaluating super-resolution results}

The primary novelty of our method lies in its ability to yield physically meaningful results, in addition to the quality performance enhancements demonstrated above using standard metrics. Given that MSE is inadequate as a surrogate for perceptual quality \citep{SaraImage2019} or physical consistency \citep{TakamotoPDEBench2022}, we supplement our analysis by computing the Euclidean norm of the gradient (Mean Squared Gradient Error, MSGE) to gauge gradient consistency, defined as:
\begin{equation}
\begin{split}
    \MSGE =& \frac{1}{N} \sum _{i=1}^N e_i e_i^T, \\
    & \text{with} \, e_i = \left[\nabla_x u_i - \nabla_x \hat{u}_i,\, \nabla_y u_i - \nabla_y \hat{u}_i \right],
\end{split}
\end{equation}
where $N$ is the number of points, $\nabla_j \hat{u}_i$ is the gradient of the approximated solution $\hat{u}$, for the real solution $u$, in direction $j = x, y$ calculated using the convolution filter at point $i$. 
 
The SSIM index is based on three components, including structure, and compares the local patterns of grid intensities of two images \citep{SaraImage2019}. Although the term ``structure" might evoke the idea of gradients, SSIM itself does not measure gradients directly. Therefore, we introduce a new index, named GSNR (Gradient Signal-to-Noise Ratio), to measure the compatibility of the approximated results' gradients with the ground truth as: 
\begin{equation}
\GSNR = 10 \log_{10} \left(\dfrac{\MAXG}{\MSGE} \right), \quad \MAXG = \left(\frac{\MAXI_i}{h_i}\right)^2,
\end{equation}
with $\MAXG$ being the maximum possible gradient of the image, $\MAXI_j$ the maximum possible value of the solution in direction $j$, and $h_j$ the smallest distance between two grid points along direction $j$. 

\subsection{Physically-consistent error analysis in time-marching super-resolution methods}
We extend our analysis to exploit the effect of the time-marching approach on the method's accuracy. Following \citet{shalev2014understanding}, the generalisation error is defined as:
\begin{equation}
    \mathcal{E}_G = \| u-\hat{u}\|_X,
\end{equation}
where $\|\cdot \|_X$ is a norm. Since the exact solution $u$ is not available, we monitor the so-called training error (using the notations for the PDE defined in \autoref{sec:method}):
\begin{equation}\label{eq:et}
    \mathcal{E}_T := \left(\sum_{i=1}^{N} \left |\frac{\partial \hat{u}_i }{\partial t}  +\mathcal{F}(\hat{u}_i)    \right|^p\right)^{1/p}.
\end{equation}

In our approach, we approximate $\frac{\partial \hat{u}_i }{\partial t} $ using a time-marching technique with a local truncation error of $\mathcal{O}(\tau^{q})$, where $\tau$ is the time step. The spatial derivatives are determined using a convolution kernel with two errors corresponding to the equation's stiffness (eigenvalues of the spatial discretisation) and the kernel operator scaled by $D$ and $D_C$, respectively. Thus, the characteristic polynomial of the obtained system reads (for details, see~\citep{behnoudfar2022variational} and Appendix C):
\begin{equation}\label{eq:char}
\begin{aligned}
  \pi\left(z;(D, D_C)\right) & := (1 + \tau (D + D_C) \beta_s)z^s \\
  & \quad + \sum_{k=0}^{s-1} \left(\alpha_k + \tau (D + D_C) \beta_k\right)z^k \\
  & = \rho(z) + \tau (D + D_C) \sigma(z),
\end{aligned}
\end{equation}
with $s,\,\{\alpha\}_k,\,\{\beta\}_k$ are the number of steps and coefficients; the functions $\rho$ and $\sigma$ of the variable $z$ are used to collect terms in the equation. From \eqref{eq:char} and the truncation error of the time-marching technique, the error corresponding to the convolution kernel can alter the accuracy of the method as:
\begin{equation}\label{eq:char2}
\begin{aligned}
  \pi(z;(D, D_C)) & = \rho(z) + \tau (D+ D_C) \sigma(z) \\
  & = \mathcal{O}(\tau^{q}) + \tau  D_C \sigma(z), \quad z \to 0
\end{aligned}
\end{equation}
with $q$ the order of the temporal truncation error, $z$ a function of the inverse of data points $N^{-1}$ and therefore $\sigma$ depending on $N^{-1}$. Then, we define another error term representing the discretisation error bounded from below as:
\begin{equation}\label{eq:err_d}
\mathcal{E}_D:= \mathcal{O}(\tau^{q}) +\tau  D_C \sigma(z).
\end{equation}

Finally, following~\citep{mishra2023estimates} and from~\eqref{eq:err_d}, we propose
\begin{equation} \label{eq:EA1}
    \mathcal{E}_G \leq C \mathcal{E}_T + C \mathcal{E}_D = C\left( \mathcal{E}_T + \mathcal{O}(\tau^{q}) + \tau  D_C \sigma(N^{-1}) \right), 
\end{equation}
where $C$ is a constant independent of the given problem's data and discretisation. 

\begin{remark}
In CN or other multi-stage approaches (e.g., RK), we have $\sigma(z) = z^q + \cdots + 1$. However, for BDF, $\sigma(z) = z^q$. Therefore, the error committed by the convolution kernel can always be better controlled in BDF than in other multistage approaches. That is, with sufficient data points resulting in $z\to0$, we have
\begin{equation} \label{eq:EA2}
\begin{aligned}
    \mathcal{E}_G &\leq C\left( \mathcal{E}_T+ \mathcal{O}(\tau^{q}) \right), \quad &&\text{for CN},\\
        \mathcal{E}_G &\leq C\left( \mathcal{E}_T+ \mathcal{O}(\tau^{q}) +\tau  D_C )\right), \quad &&\text{for RK}.
\end{aligned}
\end{equation}

\end{remark}

\subsection{Robustness and data efficiency analysis of PC-SRGAN}
We investigate the robustness of our proposed PC-SRGAN approach through several ablation experiments. For these investigations, we fix the physical model to the Allen-Cahn system with Periodic boundary conditions and employ the BDF time integrator.

\subsubsection{PC-SRGAN is data-efficient}
\begin{figure}[t!]
    \centering
    \includegraphics[width=0.85\columnwidth]{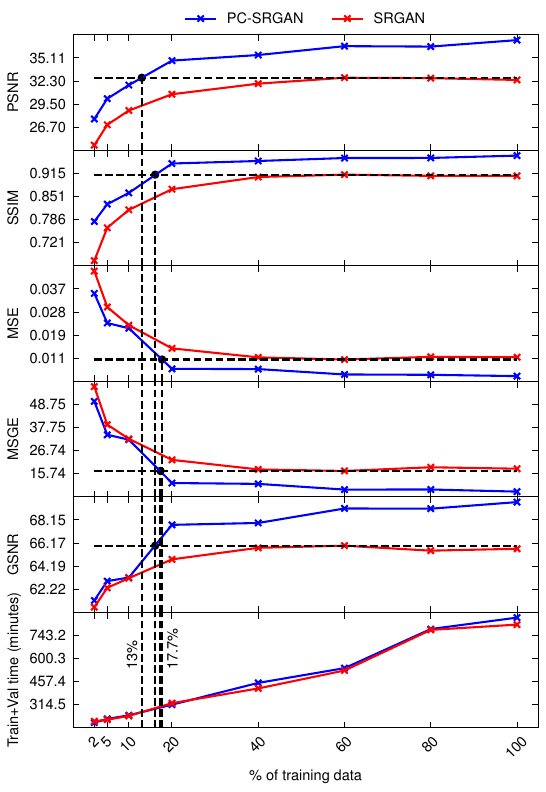}
    \caption{Study to compare the impact of the training dataset size on the quality of the results, through five performance metrics, for PC-SRGAN and traditional SRGAN. The validation set size is kept fixed across different training set sizes. PC-SRGAN surpasses the performance of SRGAN with only 13 to 17.7\% of training data based on the evaluation metrics. The difference in time costs between PC-SRGAN and SRGAN is also minimal.
    }
    \label{fig:ablation_data}
\end{figure}

We vary the amount of training data. Both PC-SRGAN and SRGAN are trained for 100 epochs using subsets randomly sampled from the complete training dataset. Specifically, the $n$\% subset refers to $n$\% of the training time series, each consisting of 100 or 50 time steps. To ensure a fair comparison, the sizes of the validation and test sets remain constant across different subsets. \autoref{fig:ablation_data} presents the performance comparison across eight different sizes of training data.

Remarkably, PC-SRGAN achieves superior performance compared to SRGAN with a small fraction of the training dataset. Just 13\% of the training data is sufficient to surpass SRGAN in PSNR, and 17.7\% for MSE; similar trends are observed for SSIM and GSNR.

Both PC-SRGAN and SRGAN exhibit similar training times on different training sets, as confirmed through 100 epochs of training and validation with the Allen-Cahn $8\times$ dataset (\autoref{fig:ablation_data}). This means we can obtain better results with PC-SRGAN than SRGAN for the same training time, or use less than 20\% of the training dataset to obtain results of the same quality with a speed-up factor of more than $3\times$.

\subsubsection{Physics-based terms have the highest contribution}
Contributions of different loss components in PC-SRGAN are examined in \autoref{tab:ablation-weight}. The loss weights were manually selected to maintain similar orders of magnitude but could be further optimised following~\citep{BasirPhysics2022}. Note that the absolute values of the loss weights do not reflect their contributions; only their relative changes do. The conventional pixel loss shows the least contribution; removing it results in a slight decrease in PSNR but unexpectedly improves SSIM, MSE, and MSGE. In contrast, the proposed physics-based loss for inner pixels has the highest impact: without it, performance degrades significantly, with PSNR dropping from 37.2748 to 34.6566. Removing the physics-based loss for boundary pixels also leads to consistent performance degradation across all four metrics.

\begin{table}[t!]
    \centering
    \caption{Ablation experiment by varying the weight coefficient of PC-SRGAN loss components, conducted on Allen-Cahn $8\times$ system with Periodic boundary conditions and BDF time integrator. Numbers in \textbf{bold} and \textit{italics} indicate the best and the second-best scores, respectively.}
    \label{tab:ablation-weight}
    \resizebox{\columnwidth}{!}{%
    \begin{threeparttable}
    \begin{tabular}{@{}*9c@{}} \toprule
        $w_1$ & $w_2$ & $w_3$ & $w_4$ & $w_5$ & PSNR $\uparrow$ & SSIM $\uparrow$ & MSE $\downarrow$ & MSGE $\downarrow$ \\ \midrule
        1.0 & 1.0 & $1\e{-3}$ & $1\e{-8}$ & 5 & \textbf{37.2748} & \textit{0.9652} & 0.0041 & 7.2372 \\
        1.0 & 1.0 & $1\e{-3}$ & $1\e{-5}$ & 5 & 30.3532 & 0.8781  & 0.0063 & 9.2439 \\
        1.0 & 1.0 & $1\e{-3}$ & $1\e{-8}$ & 1 & 37.0164 & 0.9644 & 0.0042 & 7.3940 \\
        1.0 & 1.0 & $1\e{-3}$ & $1\e{-8}$ & 0 & 36.7473 &  0.9633 & \textit{0.0040} & \textit{7.0540} \\
        1.0 & 1.0 & $1\e{-3}$ & 0 & 5 & 34.6566 &  0.9464 & 0.0053 & 9.6884 \\
        0 & 1.0 & $1\e{-3}$ & $1\e{-8}$ & 5 & 36.8332 &  \textbf{0.9653} & \textbf{0.0038} & \textbf{6.9865} \\
        1.0 & 0 & $1\e{-3}$ & $1\e{-8}$ & 5 & 35.9596 & 0.9602 & 0.0044 & 7.7241 \\
        1.0 & 1.0 & 0 & $1\e{-8}$ & 5 & 35.6935 &  0.9578 & 0.0046 & 8.2766 \\ \bottomrule
    \end{tabular}
    \begin{tablenotes}
        \item $w_1$: pixel loss, $w_2$: content loss, $w_3$: adversarial loss, $w_4$: physics-inner loss and $w_5$: physics-boundary loss
    \end{tablenotes}
    \end{threeparttable}%
    }
\end{table}

The \autoref{tab:ablation-weight} further reports the performance under different weights for the physics-based losses. Increasing the physics-inner loss weight from $1\e{-8}$ to $1\e{-5}$ leads to the worst performance across all four metrics. In contrast, reducing the physics-boundary loss weight from 5 to 1 causes only a slight performance drop. Interestingly, setting the boundary weight to zero improves MSE and MSGE, although it degrades PSNR and SSIM. Since the number of boundary pixels is much smaller than that of inner pixels, quantitative evaluation metrics may not sufficiently reflect the impact of the boundary loss. Nevertheless, its contribution is evident in the improved boundary quality of the super-resolved images.

\section{Discussion}
\begin{figure}[t!]
    \centering
    \includegraphics[width=1\linewidth]{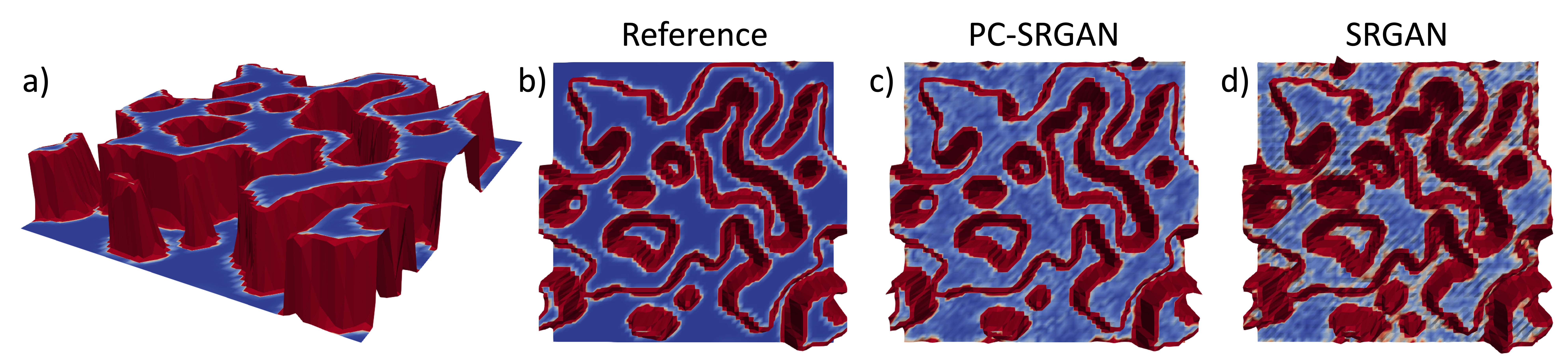}
    \caption{Comparison of Allen-Cahn $8\times$ solution distributions. (a) Results from FEM simulation (ground truth), with 3D warping by solution values and colouring by gradient magnitude. (b) Corresponding top view of ground truth image coloured by gradient magnitude. (c) Corresponding PC-SRGAN results. (d) Corresponding SRGAN results. For all subfigures, the gradient colour bar is capped from 0 (\textcolor{blue}{blue}) to 0.05 (\textcolor{red}{red}) instead of its natural maximum of ${\sim}1.15$. The 3D visualisation of the surface as a height field reveals extra information in the near-constant value areas through shading.}
    \label{fig:H1_paraview}
\end{figure}

Beyond the considerable overall improvements in quality metrics, PC-SRGAN is particularly well-suited to improve results for numerical solutions with sharp transitions and/or constant areas.
Those benefits of PC-SRGAN are particularly visible in the case of Allen-Cahn simulations, where the solution lies within the range of $[-1,1]$ and typically exhibits characteristics akin to a two-state solution, with sharp transitions between vast areas where the values are either $1$ or $-1$ (see \autoref{fig:H1_paraview}a). One of the notable contributions of PC-SRGAN is its ability to mitigate ringing artifacts, which are spurious signals near sharp transitions, commonly observed in SR outputs. Furthermore, PC-SRGAN also effectively attenuates the checkerboard patterns often present in regions with nearly constant values. Visualising the solution using gradient magnitude colouring, with the colour bar capped at low values, highlights the quality of gradients in these zones (\autoref{fig:H1_paraview}(b-d)). Beyond the quality of the transitions and the constant areas, the improvement of PC-SRGAN over SRGAN is also evident in the removal of larger artificial jumps in the solution, contributing to the quantitative enhancement of gradient consistency, as demonstrated in \autoref{tab:results}.

The advantages of our approach also extend beyond SR imaging and into the subsequent usage of those results. For instance, when machine learning is employed to produce surrogate models for numerical simulations, a high image quality measured solely by pointwise MSE, a \emph{first-order} metric, may conceal poor-quality results concerning downstream tasks. In numerical modelling, one often seeks accuracy in the gradients of model quantities on top of solution accuracy, as resulting forces (e.g., diffusion) are commonly derived from gradients, and the smoothness of the solution plays a crucial role in the quality of the results. The additional image quality assessment metric we introduced, GSNR, intuitively captures the compatibility of approximated gradients with ground truth; it helps demonstrate how the added physical consistency outperforms a standard pointwise loss approach.

\begin{figure}[t!]
    \centering
    \includegraphics[width=1\linewidth]{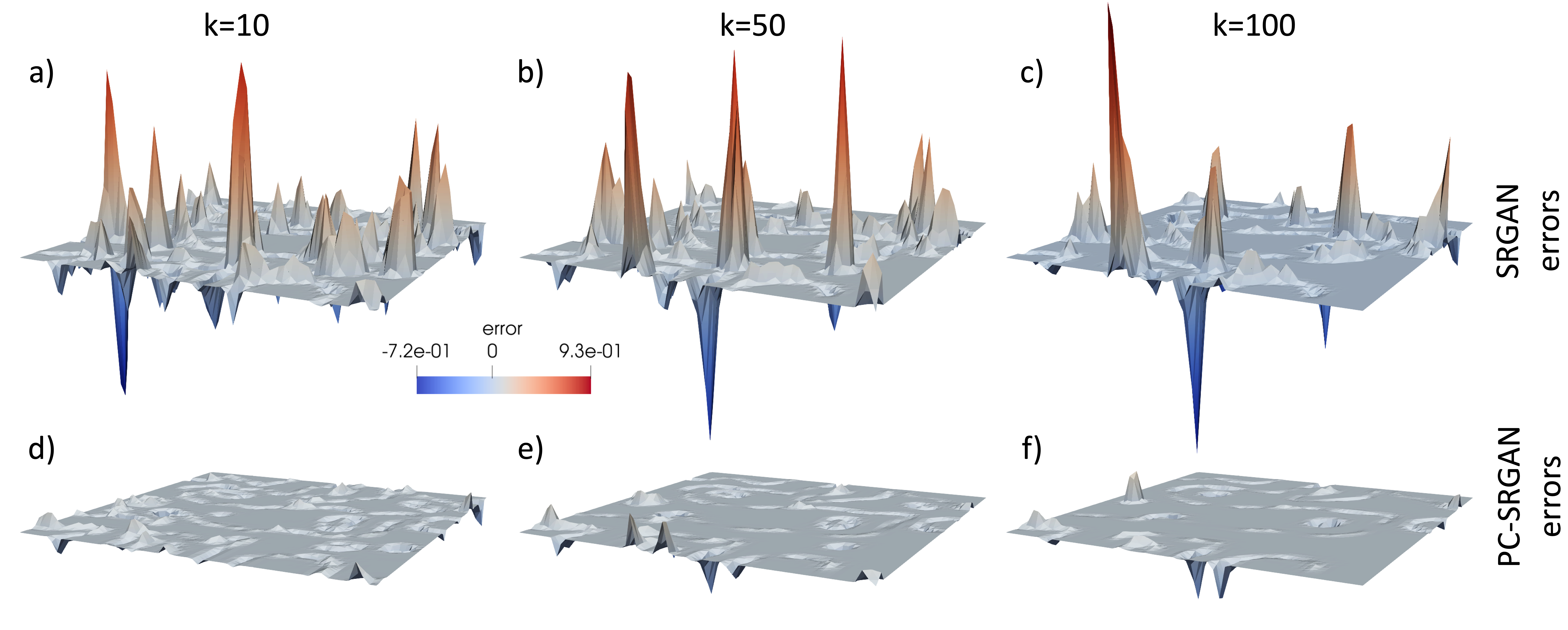}
    \caption{
    Comparison of PC-SRGAN's and SRGAN's surrogate capabilities. Both methods start from low-resolution numerical simulation results to generate high-resolution images, from which simulations are restarted and then continued. Those numerical results are compared with FEM results run on the fine mesh from the beginning, and the errors are plotted for SRGAN (a-c) and PC-SRGAN (d-f) at time steps $k=10$ (a, d), $k=50$ (b, e) and $k=100$ (c, f) after the restart. The errors with PC-SRGAN are at least one order of magnitude smaller than with SRGAN, with the discrepancies growing over time.
    }
    \label{fig:FEM_errors}
\end{figure}

The PC-SRGAN results can be used as a physically-consistent surrogate, in the sense that they produce valid initial solutions for further simulations. Starting from coarse-grid solutions of a finite element solver run for 150 time steps on an $8 \times 8$ mesh, we obtain the corresponding $64 \times 64$ higher-resolution solution using either SRGAN or PC-SRGAN. Those two finer-grid solutions are then taken as inputs for the FEM solver to continue marching in time for another 100 time steps, and the two resulting time series are compared with the reference obtained by FEM from the initial time step for that finer mesh. Although additional metrics could be developed to evaluate surrogate capability, the results (\autoref{fig:FEM_errors}) illustrate the impact of physically consistent solutions on long-term FEM computations and showcase advantageous features of our approach. The error derived from PC-SRGAN images is an order of magnitude smaller than that derived from SRGAN images, where growing instabilities push the solution much faster towards a different state. In that respect, SR can contribute to a multi-fidelity strategy \citep{Chen2024} and process understanding.

For time series prediction, in contrast to some studies in this domain, we explicitly consider the principled choice of time integration strategies in the design of loss functions. Our results demonstrate that this choice is consequential. The assumption of a constant time step ($\tau = 1$) in many machine learning studies should certainly be revisited, and a proper time analysis considered to admit a consistent physical model, as shown in our derived error estimation~\eqref{eq:EA1} and \eqref{eq:EA2}. Moreover, our method employing a rigorous time integration technique scales temporal and spatial operators effectively by its structure, ensuring causality and precise temporal estimations with respect to the truncation error. While the \emph{pushforward trick} \citep{BrandstetterMessage2022}, known for enhancing the learning of time-correlated PDEs \citep{TakamotoPDEBench2022}, attempts to learn this scaling, the incurred error in the temporal domain remains unclear. Besides, unlike the pushforward approach, the stability of our method is inherently derived from its structure, obviating the necessity for an additional loss term to maintain stability.

Our approach exhibits versatility in handling a wide range of physical systems, as demonstrated through various examples with different boundary conditions. This adaptability is showcased by the ability of PC-SRGAN to accommodate diverse simulation scenarios, which makes it suitable for applications across different scientific domains. Our framework is designed to learn both the temporal dynamics and spatial behaviour of the physical system, including cases where the equations might be partially unknown, as long as the time derivative is of first order, such as parabolic and elliptic PDEs. However, it could be extended to encompass hyperbolic systems with second-order derivatives in time. Such expansion would broaden the scope of applications and allow for the investigation of phenomena characterised by wave-like behaviour and transient dynamics. 

In conclusion, this study illustrates the transformative effect that machine learning can bring to simulation~\citep{Azizzadenesheli2024}. Beyond the significant improvements from an image processing perspective, both in terms of quantitative output quality and numerical efficiency with reduced training costs, the enhanced physical realism of images produced by PC-SRGAN marks a significant advancement for domain experts. Enhancing physical consistency in SR approaches improves the quality, interpretability and reusability of reconstructed images \citep{ShenDifferentiable2023} and facilitates more accurate and insightful analyses of transient numerical simulation results. The integration of physical modelling and machine learning in PC-SRGAN demonstrates the potential of broader impact across diverse fields of science and engineering.

\appendices

\section{Model problem and notation}
\label{sec:model_and_notation}
For an open bounded domain $\Omega$ with Lipschitz boundary $\partial \Omega$, outward unit normal vector $\vn$, the transient reactive flow problem defines as the following convection-diffusion-reaction Initial-Boundary-Value Problem (IBVP):
\begin{equation}\label{eq:prob_app}  
\frac{\partial u }{\partial t}  - \Nabla \cdot \DD \Nabla +
 \bb\cdot \Nabla u + R(u) =f,\quad \text{ in } \Omega_T.  
\end{equation}
where $\DD\in\SLOinf$ denotes the diffusion tensor; $\bb\in [L^2(\Omega)]^N$ the advection field; $f\in\SLTO$ is the source function; and $R(u)$ represents the nonlinear reaction rate.
Note that the gradient operator $\Nabla$ refers to the spatial gradient operator, e.g., $\Nabla (\cdot) = \{\frac{\partial (\cdot) }{\partial x},\frac{\partial (\cdot) }{\partial y}  \}^{\text{T}}$. Initial and boundary data are given such that~\eqref{eq:prob_app} admits a well-posed equation.

We demonstrate the effectiveness of our proposed approach for IBVP problems, and extend it to different boundary conditions: Dirichlet, Neumann and Periodic. 

\subsection{Allen-Cahn equation}

The Allen-Cahn equation is a non-linear reaction-diffusion equation of IBVP type~\eqref{eq:prob_app}, which models the phase separation in multi-component alloy systems. It has real-world applications in a vast number of areas, such as plasma physics, biology, quantum mechanics \citep{khater2020analytical}, microstructure evolution~\citep{chen2002phase}, image segmentation~\citep{benevs2004geometrical}, biological pattern formation~\citep{inan2020analytical}, and nonlocal buckling~\citep{behnoudfar2022localized}. The \emph{strong form Allen-Cahn equations} finds the phase concentration scalar field $\varphi \left( \boldsymbol{x} , t \right) $ that fulfils
\begin{equation}
\begin{array}{r l l}
\frac{\partial \varphi }{\partial t}  + M_{\eta} \left( \dfrac{\Phi' \left(\varphi\right)}{\varepsilon^2} - \nabla \cdot \nabla \varphi \right) &= 0 & \text{in} \ \ \Omega_T, 
\end{array}
\label{eq:AllenCahn}
\end{equation}
where $M_{\eta}$ is a constant mobility parameter, $\Phi'$ is the derivative of the Helmholtz free energy density with respect to $\varphi$, and $\varepsilon$ is a small positive constant called the gradient energy coefficient, related to the interfacial energy. We consider two cases:
The first one has Neumann boundary ($\nabla \varphi \cdot \mathbf{n}  =0 $) on $\partial \Omega$ with respect to its normal vector $\mathbf{n}$, and in the second, we impose Periodic boundary conditions on the top and right-hand sides of the boundary.

\begin{remark} Here, we use the following Helmholtz free energy density:
\begin{align}
\begin{split}
\Phi \bigl(\varphi\bigr)& = \frac{1}{2 T } \bigl(\bigl( 1 + \varphi \bigr) \ln\bigl( 1 + \varphi \bigr)\\
&+ \bigl( 1 - \varphi \bigr)\ln\bigl( 1 - \varphi \bigr)\bigr) - \frac{1}{2} T_c \varphi^{2},
\end{split}
\end{align}
where $T $ is a relative temperature and $T_c$ a critical temperature. The gradient energy coefficient is computed as
\begin{equation}
\varepsilon = \frac{h m}{2 \sqrt{2} \tanh^{-1} \left(0.9 \right)},
\end{equation}
where we set $h=1/64$ and $m=1$ as an arbitrary natural number \citep{LI20101591}.
\end{remark}

\subsection{Eriksson-Johnson problem}
We consider a well-known example of transient convection-diffusion-reaction, the Eriksson-Johnson problem \citep{eriksson1993adaptive}. Example real-world applications of this problem include computational mechanics \citep{johnson1992adaptive}, fluid flow, diffusion and elasticity \citep{pozrikidis1998computational}. This problem still has IBVP form~\eqref{eq:prob_app}, specifically
\begin{equation}\label{eq:conv_diff_eriksson}  
\frac{\partial u }{\partial t}  - \DD \, \Delta u +
 \bb\cdot \Nabla u + K u\left(u-1\right)=0,\quad \text{ in } \Omega_T.  
\end{equation}
The initial condition reads:
\begin{equation}
u(x,y,0) = \left( e^{ \lambda_1 \, x}\, - e^{\lambda_2 \, x}\,  \right) +\cos (\pi \, y)\,  \frac{ e^{ \delta_2 \, x}\, - e^{ \delta_1 \, x} }{e^{ - \delta_2 }\, - e^{ - \delta_1} },
\end{equation} 
and the time-dependent Dirichlet boundary conditions on $\partial\Omega$ are: 
\begin{equation}
u(x,y,t) = e^{-t} \left( e^{ \lambda_1 \, x}\, - e^{\lambda_2 \, x}\,  \right) + \cos (\pi \, y)\,  \frac{ e^{ \delta_2 \, x}\, - e^{ \delta_1 \, x} }{e^{ - \delta_2 }\, - e^{ - \delta_1} },
\end{equation} 
where
\begin{equation} 
 \lambda_{1,2} =  \frac{ - 1 \pm \sqrt{1 - 4 \, \epsilon \, l}}{ -2 \, \epsilon },  \qquad 
 \delta_{1,2} =  \frac{  1 \pm  \sqrt{1 + 4 \, \pi^2 \, \epsilon^2}}{ 2 \, \epsilon }.  
\end{equation}
The problem domain is $\Omega_T = ( -1, 0 ) \times ( -0.5,0.5  ) \times ( 0 ,0.5 ) $.

\subsection{Discretisation}
We use a Galerkin Finite Element Method (FEM) to obtain the semi-discrete space that represents PDEs, simulating them in the space-time domain~(for details, see \citep{hughes2012finite}). Next, we exploit the generalised-$\alpha$ method to approximate the temporal operator~\citep{jansen2000generalized}. We gathered the results as a benchmark for further studies, which is released publicly at \citep{hasan2025transient}. 

For super-resolution model development and evaluation, each time series in the dataset is a simulation of 100 time steps. From the whole dataset, we randomly pick 15\% of the time series (of which 10\% time series includes the whole 100 time steps and 5\% time series includes the last 50 steps of the simulation) as the validation set and then similarly another 15\% as the hold-out test set. The remaining data are used for training. The validation set is used to validate models during training, and we use the hold-out test set only once during our final testing of the models. \autoref{tab:dataset} reports the number of samples in training, validation and test splits across four datasets used in this study.

\begin{table*}[!t]
    \centering
    \caption{Statistics of the datasets, showing the number of total time steps (snapshots) used for training, validation, and testing of PC-SRGAN for different problems with specific boundary conditions.}
    \label{tab:dataset}
    \begin{tabular}{c*2l*4c}
    \toprule
        \textbf{Scale factor} & \textbf{Problem} & \textbf{Boundary} & \textbf{Train} & \textbf{Validation} & \textbf{Test} & \textbf{Total} \\
    \midrule
        $8\times$ & Allen-Cahn & Periodic & 55,062 & 9,065 & 7,973 & 72,100 \\
                    &   & Neumann & 15,732 & 2,590 & 2,278 & 20,600 \\
    \cmidrule{2-7}
        &   Eriksson-Johnson & Dirichlet & 51,875 & 8,494 & 7,352 & 67,721 \\ \midrule
        $4\times$ & Allen-Cahn & Periodic & 27,662 & 4,506 & 3,882 & 36,050 \\ 
    \bottomrule
    \end{tabular}
\end{table*}

\section{Convolution filters for calculating spatial operators}
\label{app:conv}
The Laplacian is computed through the convolution with a filtering kernel
\[
\alpha
\begin{bmatrix}
    0 & 1 & 0 \\
    1 & 4 & 1 \\
    0 & 1 & 0 \\
\end{bmatrix},
\]
where the multiplying coefficient chosen as $\alpha=9.894$ for grid of $64\times64$. The coefficient is obtained by matching the results using an analytical solution $f(x,y)=\cos(n\pi x) \sin(m\pi y)$ with $n,\,m \in \mathbb{R}$ showing $C^\infty$ regularity at the given resolution.
Also, for directional derivatives, we have
\begin{equation}
   \beta  \begin{bmatrix}
    1 & 0 & -1 \\
    3.5887  & 0 &  -3.5887\\
   1 & 0 & -1 \\
\end{bmatrix},
\end{equation}
in x-direction, and
\begin{equation}
    \beta \begin{bmatrix}
    1 & 3.5887 & 1 \\
    0 & 0 & 0\\
   -1 & -3.5887 & -1 \\
\end{bmatrix},
\end{equation}
in y-direction with the empirical value $\beta= - 5.645$.

\section{Time integrators}
\label{app:time}
We use two classes of time integrators in this paper.

\subsection{Multistep schemes}
In general, multistep methods use data from the previous $s$ steps to estimate the solution at the next time step. Let $\mathcal{F}_{n}$ approximates $\mathcal{F}(u(t_n))$. Then, a general linear multistep method discretises our problem in time as ~\citet{butcher2016numerical}:
\begin{equation*}
    \begin{split}
        U_{n+s} &+ \alpha_{s-1} \cdot U_{n+s-1} + \alpha_{s-2} \cdot U_{n+s-2} + \cdots + \alpha_{0} \cdot U_{n} \\
        & + \tau \left( \beta_s \cdot \mathcal{F}_{n+s} + \beta_{s-1} \cdot \mathcal{F}_{n+s-1} + \cdots + \beta_0 \cdot \mathcal{F}_{n} \right) = 0,
    \end{split}
\end{equation*}
where different coefficient choices for $\alpha_i$ and $\beta_i$ result in different methods. Backward Differentiation Formula (BDF) methods are a widely used class of multistep techniques with the general form~\citep{ascher1998computer,iserles2009first}:
\begin{equation}\label{eq:bdf}
 \tau\beta \mathcal{F}_{n+s} + \sum_{j=0}^{s}\alpha_j U_{n+j}=0,
\end{equation}
where $\alpha_j$ and $\beta$ allow the method to reach the accuracy of order $s$. We refer the reader for the details on the determination of the coefficient to \citet{ascher1998computer,iserles2009first}.

\subsection{Characteristic polynomial}
A linear multistep or multistage method exploits a characteristic polynomial to analyse the method's stability and accuracy.
A general form of the characteristic polynomial with a time-step size $\tau$ reads~\citep{suli2003introduction}:
\begin{equation}\label{eq:poly}
  \begin{split}
    \pi(z;\delta) &= (1+\tau \delta \beta_s)z^s + \sum_{k=0}^{s-1} \left( \alpha_k + \tau \delta \beta_k \right)z^k \\
    &= \rho(z) + \tau \delta \sigma(z),
  \end{split}
\end{equation}

with $\delta$ representing error due to approximating the spatial operator~$\mathcal{F}(\cdot)$.
A scheme is consistent if the local truncation error goes to zero faster than the time step $\tau$ as $\tau \to 0$. Also, for problems with first-order derivative in time, we have~\citep{behnoudfar2022variational}:
\begin{equation}
    \rho(\exp(\tau))+\tau \delta \sigma\left(\exp(\tau)\right) = \mathcal{O}\left(\tau ^{s+1} \right),
\end{equation}
with $\rho$ and $\sigma$ being the collected terms in~\eqref{eq:poly}. 

\section{Architecture and training details}
\label{app:network}
\subsection{Architecture details}
The proposed PC-SRGAN architecture is based on the SRGAN architecture \citep{ledig2017photo}. We used a public implementation of SRGAN available on GitHub\footnote{\url{https://github.com/Lornatang/SRGAN-PyTorch} (Commit: \href{https://github.com/Lornatang/SRGAN-PyTorch/tree/cf1309568f9f9733a1ca2f5a798e6ae909dddd62}{cf13095})}. Apart from the necessary adaptations required to solve the problem at hand, all hyperparameters, including the loss weights and architecture, remained unchanged. The necessary adaptations include an 8$\times$ upscaling factor instead of 4$\times$ for our $8\times$ super-resolution experiments and a different input resolution.

As with other GANs, SRGAN \citep{ledig2017photo} consists of two major networks: a generator and a discriminator. In SRGAN, the generator $G$ adopts a deep Residual Network (ResNet) architecture~\citep{he2015deep}. It begins with a convolutional layer ($9\times9$ kernel) and PReLU activation to extract low-frequency information, followed by a series of residual blocks with $3\times3$ convolutions, batch normalisation, and PReLU to capture high-frequency details. A subsequent convolutional layer fuses the low- and high-frequency information. Resolution is increased through upsampling blocks using convolution, pixel-shuffle operations, and PReLU activations. Each upsampling block upscales by a factor of 2 using a 3$\times$3 kernel. The desired upscaling factor determines the number of upsampling blocks. Finally, a reconstruction layer ($9\times9$ convolution) produces the high-resolution output, clamped appropriately for numerical stability based on the solution range. For example, the Allen-Cahn and Eriksson-Johnson solutions range between $[-1, +1]$ and $[0, 2]$, respectively. To ensure numerical stability with the logarithmic calculation in the loss function, the output is clamped to $[-1+10^{-2}, 1-10^{-2}]$ for the Allen-Cahn solution. 

The discriminator $D$ follows a VGG-like architecture~\citep{simonyan2015deep}, operating on $64\times64$ input images. It consists of sequential convolutional layers ($3\times3$ kernels) with batch normalisation and leaky ReLU activations, progressively increasing channels while reducing spatial dimensions. The features are flattened and passed through a fully connected layer (1024 units, leaky ReLU), followed by a final sigmoid layer to classify images as real or generated.

Our additional experiment with Enhanced SRGAN (ESRGAN) is also based on a public implementation available on GitHub\footnote{\url{https://github.com/Lornatang/ESRGAN-PyTorch} (Commit: \href{https://github.com/Lornatang/ESRGAN-PyTorch/tree/5a72867b7d344e27acd304f18727b94eb6a969ec}{5a72867})}. \citet{esrgan} proposed ESRGAN, which uses Residual-in-Residual Dense Block (RRDB) -- having multiple convolutional layers with multi-level residual connections -- as its generator network.

\subsection{Training details}
The training is performed using an adversarial approach, where the generator and discriminator models are trained alternately. During the generator training phase, the discriminator's backpropagation is disabled, and the generator is trained to minimise the combined loss function $\mathcal{L}_\text{PC-SRGAN}$ consisting of pixel loss, content loss (VGG19 feature loss), adversarial loss and physics-based losses (inner and boundary losses). In the discriminator training phase, the discriminator's backpropagation is enabled, and it is trained to correctly classify real and generated high-resolution images.

The relative importance of each loss component is controlled by tunable weights in a way to matches their magnitudes. The pixel loss and content loss are assigned equal weights of 1.0, while the adversarial loss has a lower weight of 0.001. The physics-inner and physics-boundary losses have weights of $10^{-8}$ and 5.0 for the Allen-Cahn problem and $10^{-2}$ and 100 for the Eriksson-Johnson problem.

During training, an Exponential Moving Average (EMA) of the generator weights is maintained to stabilise the training process. Both generator and discriminator models are optimised using the Adam optimiser \citep{kingma2015adam} with a learning rate of 0.001, betas of (0.9, 0.999), and a weight decay of 0.0. The learning rate is adjusted using a MultiStepLR scheduler with a decay factor of 0.5 at 9\textsuperscript{th} epoch.

The training process is performed for 100 epochs, with a batch size of 16. During training, the models are evaluated on a separate validation dataset using Peak Signal-to-Noise Ratio (PSNR), Structural Similarity Index Measure (SSIM), Mean Squared Error (MSE), and Mean Squared Gradient Error (MSGE) metrics. Checkpointing is employed to save the best and last models based on the PSNR and SSIM evaluation metrics. Once the model and hyperparameter setup are finalised for each scenario (e.g., Allen-Cahn with Periodic boundary conditions), we evaluate its performance on the hold-out test set. All images in our dataset are single-channel, and all metrics except LPIPS (Learned Perceptual Image Patch Similarity) are computed directly on this channel. Since LPIPS requires 3-channel inputs, we replicate the single channel across the RGB dimensions.

For ESRGAN, we adopt the same hyperparameters reported in their original paper \citep{esrgan}. First, RRDB is trained from scratch for a maximum of 120 epochs (i.e., 400K iterations), with the same checkpointing strategy as SRGAN and PC-SRGAN. Following \citep{esrgan}, this pre-trained RRDB model is used as the generator in ESRGAN and PC-ESRGAN, where it is fine-tuned using generative adversarial training with a discriminator network. After training a maximum of 120 epochs, the best checkpoint is evaluated for super-resolution on the same hold-out test set. For PC-ESRGAN's physics-based losses, we use the same loss weights as PC-SRGAN.


\printbibliography

\begin{IEEEbiography}[{\includegraphics[width=1in,height=1.25in,clip,keepaspectratio]{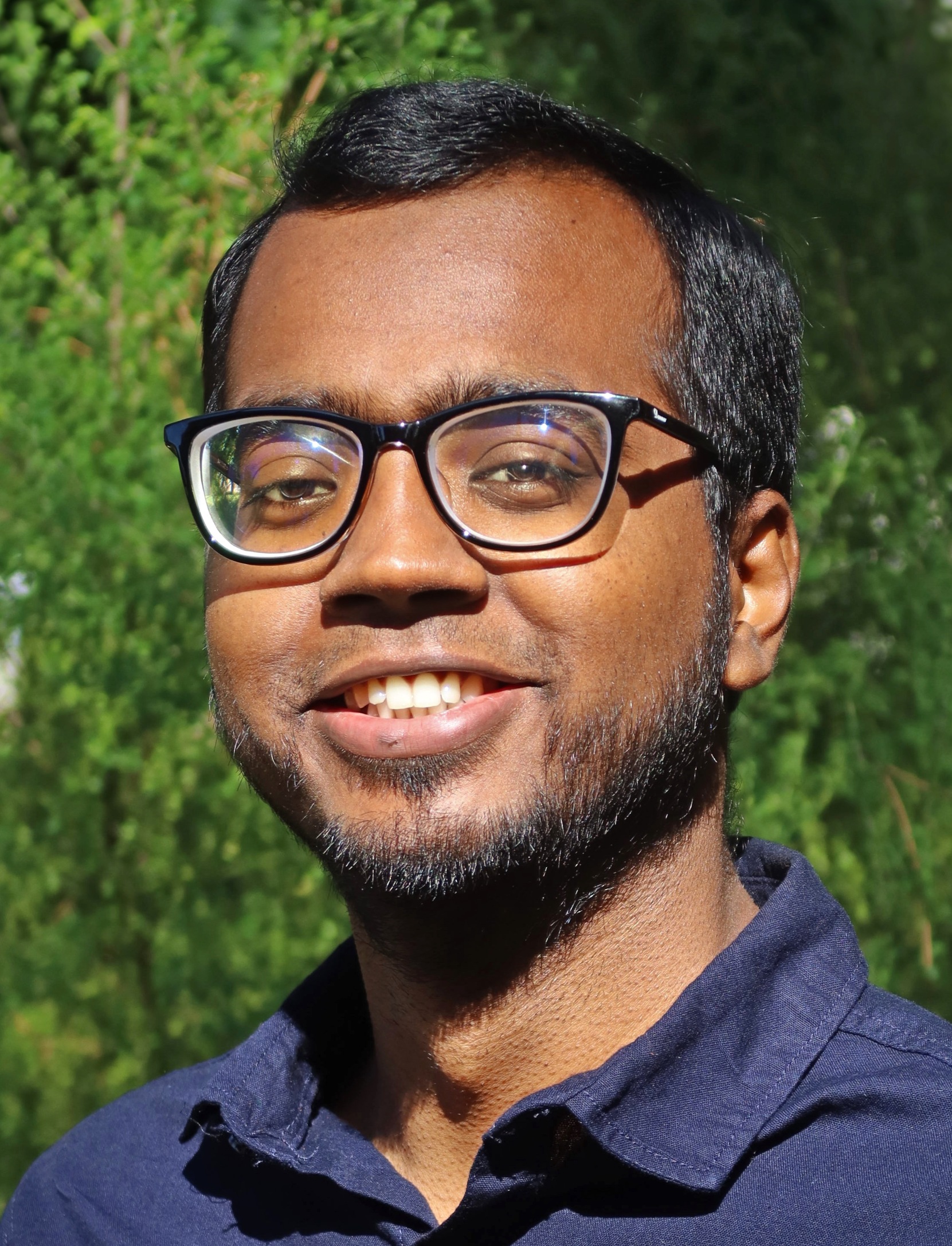}}]{Md Rakibul Hasan} received his BSc (Hons) (2019) and MSc (2021) degrees from Khulna University of Engineering \& Technology, Bangladesh. Currently, he is a PhD candidate in Computing at Curtin University, Western Australia. His overarching research interest includes advancing deep learning algorithms to solve various natural language processing and computer vision tasks. 
\end{IEEEbiography}

\begin{IEEEbiography}[{\includegraphics[width=1in,height=1.25in,clip,keepaspectratio]{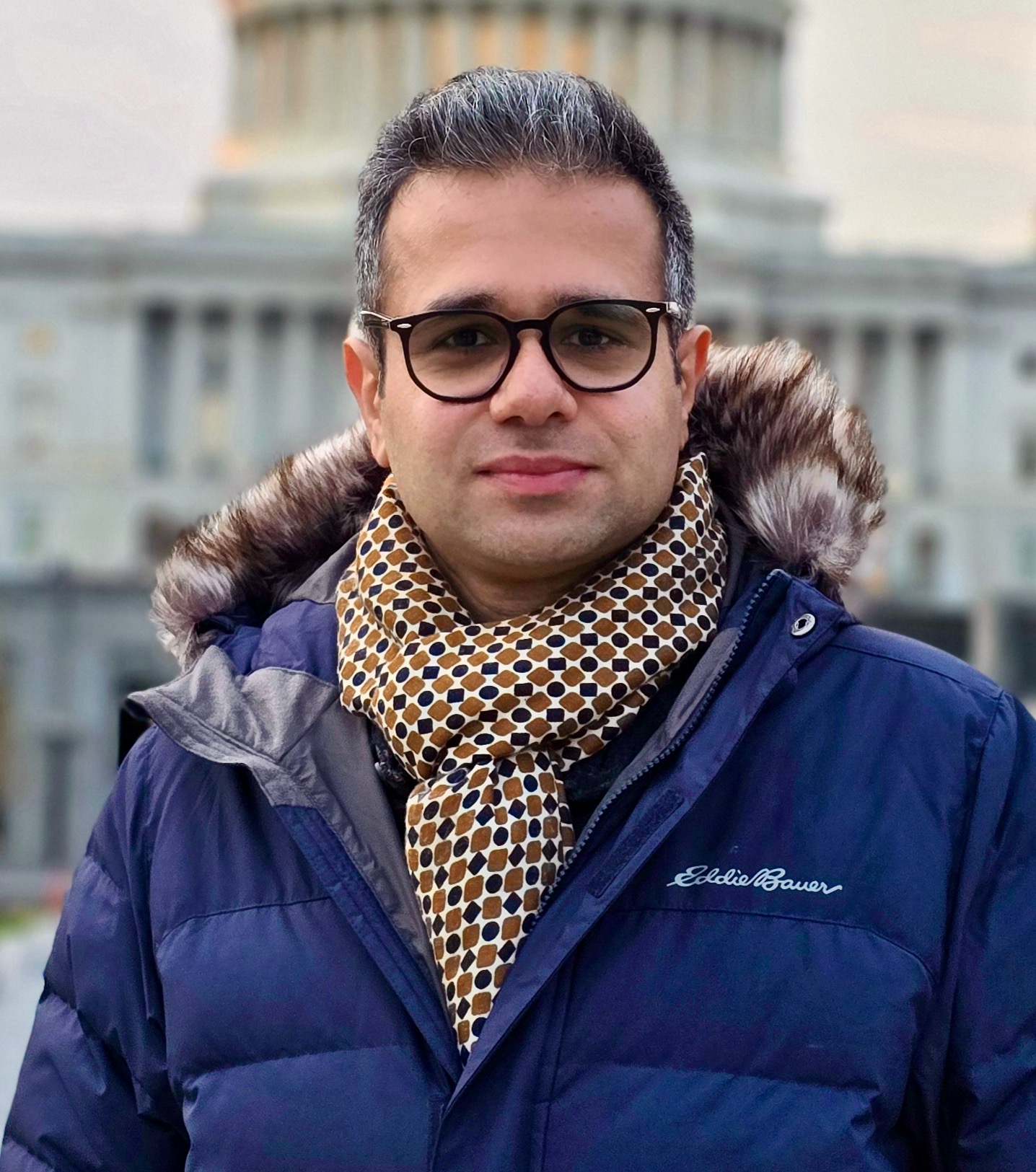}}]{Pouria Behnoudfar} is a Visiting Assistant Professor at the Department of Mathematics, University of Wisconsin–Madison. He received his Ph.D. in Earth Sciences, where he focused on numerical and physical modelling of multi-physics problems. His current research centres on physically consistent machine learning, uncertainty quantification, and data-driven modelling of dynamical systems.

\end{IEEEbiography}

\begin{IEEEbiography}[{\includegraphics[width=1in,height=1.25in,clip,keepaspectratio]{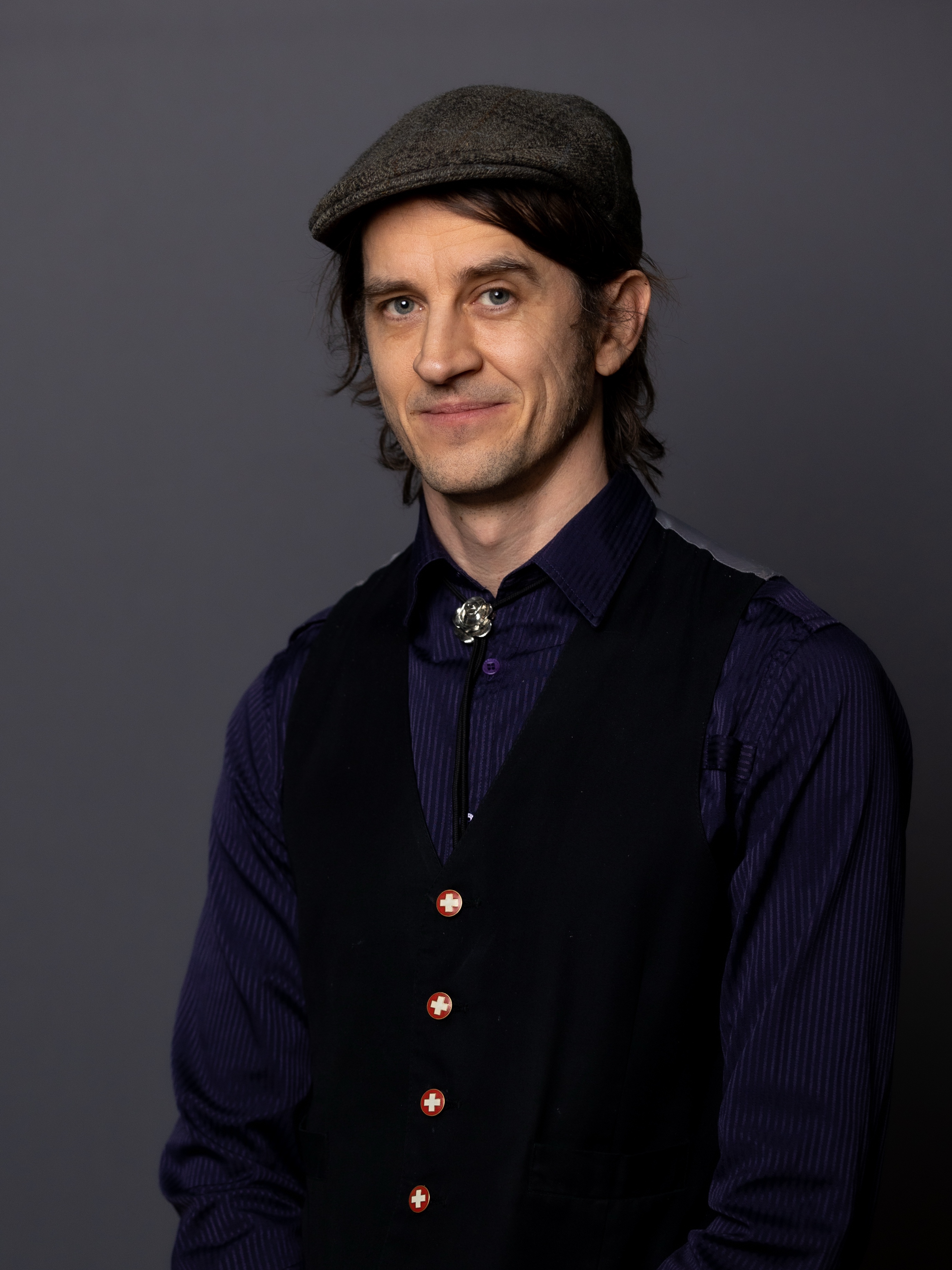}}]{Dan MacKinlay} is a statistician with CSIRO in Melbourne, Australia. His research focuses on hybrid machine learning methods for the physical sciences, including Bayesian neural networks and Gaussian processes, with applications in AI safety and scientific discovery. He received his MSc in statistics from the Swiss Federal Institute of Technology and a PhD at the University of New South Wales.

\end{IEEEbiography}

\begin{IEEEbiography}[{\includegraphics[width=1in,height=1.25in,clip,keepaspectratio]{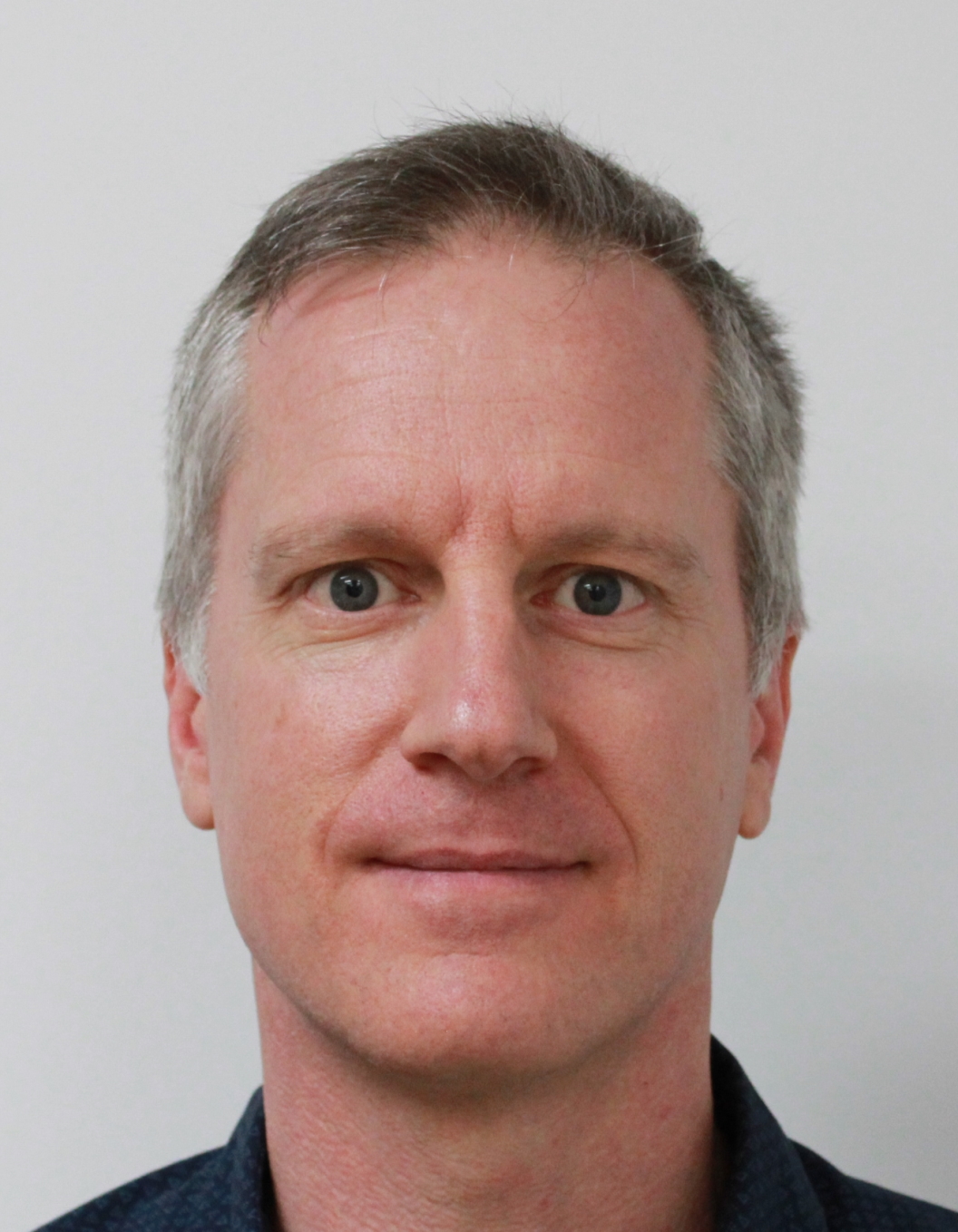}}]{Thomas Poulet} is a Research Scientist at CSIRO specialising in geomechanics, numerical modelling, and physics-based machine learning. With a PhD in geology, his work advances predictive models of geological processes through multi-physics simulations, contributing to mineral exploration, geothermal energy, and computational geoscience.
\end{IEEEbiography}

\vfill

\end{document}